\documentclass[aps,prb,preprint,superscriptaddress]{revtex4-1}

\usepackage{pdfpages}
\usepackage{graphicx}

\makeatletter
\AtBeginDocument{\let\LS@rot\@undefined}
\makeatother

\begin{document}

\title{\textit{Ab initio} dipolar electron-phonon interactions in two-dimensional materials}

\author{Tianqi Deng}
\author{Gang Wu}
\email[Corresponding author: ]{wug@ihpc.a-star.edu.sg}
\author{Wen Shi}
\author{Zicong Marvin Wong}
\affiliation{Institute of High Performance Computing, Agency for Science, Technology and Research, 1 Fusionopolis Way, \#16-16 Connexis, Singapore 138632, Singapore}
\author{Jian-Sheng Wang}
\affiliation{Department of Physics, Faculty of Science, National University of Singapore, 2 Science Drive 3, Singapore 117551, Singapore}
\author{Shuo-Wang Yang}
\email[Corresponding author: ]{yangsw@ihpc.a-star.edu.sg}
\affiliation{Institute of High Performance Computing, Agency for Science, Technology and Research, 1 Fusionopolis Way, \#16-16 Connexis, Singapore 138632, Singapore}

\date{\today}

\begin{abstract}
We develop an \textit{ab initio} formalism for dipolar electron-phonon interactions (EPI) in two-dimensional (2D) materials. Unlike purely longitudinal Fr{\"o}hlich model, we show that the out-of-plane dipoles also contribute to the long-wavelength non-analytical behavior of EPI. And the 2D dipolar EPI plays an important role not only in the typical polar material $\text{MoS}_2$, but also in graphane and fluorinated graphene. By incorporating this formalism into Wannier-Fourier interpolation, we enable accurate EPI calculations for 2D materials and subsequent intrinsic carrier mobility prediction. The results show that Fr{\"o}hlich model is inadequate for 2D materials and correct long-wavelength interaction must be included for the reliable prediction.
\end{abstract}

\maketitle


\section{Introduction}
The couplings between electrons and atomic vibrations, i.e. the electron-phonon interactions (EPIs) play major roles in many solid-state phenomena \cite{Giustino2017}, and there is continuing effort to predict the EPIs from accurate simulation approaches. \textit{Ab initio} calculation methods, particularly those based on the density functional perturbation theory (DFPT), have advanced significantly in recent years for three-dimensional (3D) materials, facilitating parameter-free simulations of charge transport \cite{Vukmirovic2012,Liu2017,Ponce2019,Shi2019,Deng2020}, heat transport \cite{Liao2015}, phonon-assisted optical absorption \cite{Noffsinger2012}, superconductivity \cite{Margine2014,Heil2017}, and polaron formation \cite{Verdi2017,Sio2019a,Sio2019}, to name a few. In these calculations, dense Brillouin zone sampling are required especially in the presence of long-range interaction where EPI varies rapidly in the Brillouin zone. Direct DFPT calculations become expensive and may not be feasible in such cases, where interpolation methods including linear or Wannier-Fourier interpolations are needed to reduce the computational cost. To achieve accurate and efficient interpolations, correct long-wavelength limit of EPI is particularly important. However, the reduced dimensionality poses additional challenge for two-dimensional (2D) materials due to different long-wavelength Coulomb interaction \cite{Huser2013,Qiu2016,Rasmussen2016}. Particularly, the long-wavelength EPI in 2D polar materials converges to a finite value \cite{Mori1989,Rucker1992}, in contrast to the 3D ${\left|{\mathbf{q}}\right|^{-1}}$ divergence for small phonon momentum ${\mathbf{q}}$ \cite{Frohlich1954}. Although in 2D materials the EPI does not diverge, it presents a sharp cusp around ${\mathbf{q}}\to{\mathbf{0}}$. Such cusp slows down the Brillouin zone integration and is difficult to approximate using Fourier or Wannier interpolation. Therefore, simple Wannier-Fourier interpolation or previously developed 3D Fr{\"o}hlich models \cite{Ponce2015,Sjakste2015,Verdi2015} lead to incorrect long-wavelength EPI and results in reduced predictive power for 2D materials \cite{Li2019,Ma2020}. Consequently, special care should be taken for EPI calculations in 2D polar materials \cite{Ponce2020}, and it takes expensive DFPT calculations on a dense $\mathbf{q}$-grid to achieve converged prediction \cite{Li2015,Sohier2018}.

Previously, first-principles-based models have been proposed for such interactions between electrons and longitudinal optical (LO) phonons \cite{Kaasbjerg2012,Sohier2016,Zhou2020}, and Sohier et al\cite{Sohier2016} proposed a 2D variation of the 3D Fr{\"o}hlich interaction which achieved agreement with finite ${\mathbf{q}}$ DFPT results. However, these materials are not strictly 2D objects, and their out-of-plane vibration and distribution should be considered if the our-of-plane  Born effective charge is significant. The dipolar potentials in earlier models were either purely 2D or depended on a thickness parameter \cite{Kaasbjerg2012,Sohier2016,Zhou2020}. Moreover, previous models only addressed the longitudinal dipoles from LO phonons and cannot explain the non-analyticity observed in transverse phonon modes, specifically the purely out-of-plane $A_{1}^{'}$ homopolar phonon in transition metal dichalcogenides (TMDCs). Therefore, the absence of a generally applicable model for dipolar EPI is still a major obstacle that prevents accurate and accelerated EPI simulations for 2D materials.

To address the abovementioned issues, we propose an \textit{ab initio} formalism for dipolar EPI in 2D materials. In this work, we show that in 2D materials, both in-plane longitudinal dipoles and out-of-plane transverse dipoles contribute to the non-analytical long-wavelength EPI. This contrasts with 3D or 2D Fr{\"o}hlich models and emphasizes the quasi-2D nature. The interaction is present in both typical polar materials like $\text{MoS}_2$, and materials with covalent bond such as hydrogenated graphene (graphane). It is also incorporated into Wannier-Fourier interpolation  \cite{Giustino2007,Ponce2016} to accelerate EPI calculation for 2D materials, which are demonstrated by phonon-limited Boltzmann transport equation calculations.

\begin{figure}
\includegraphics[width=8.6cm]{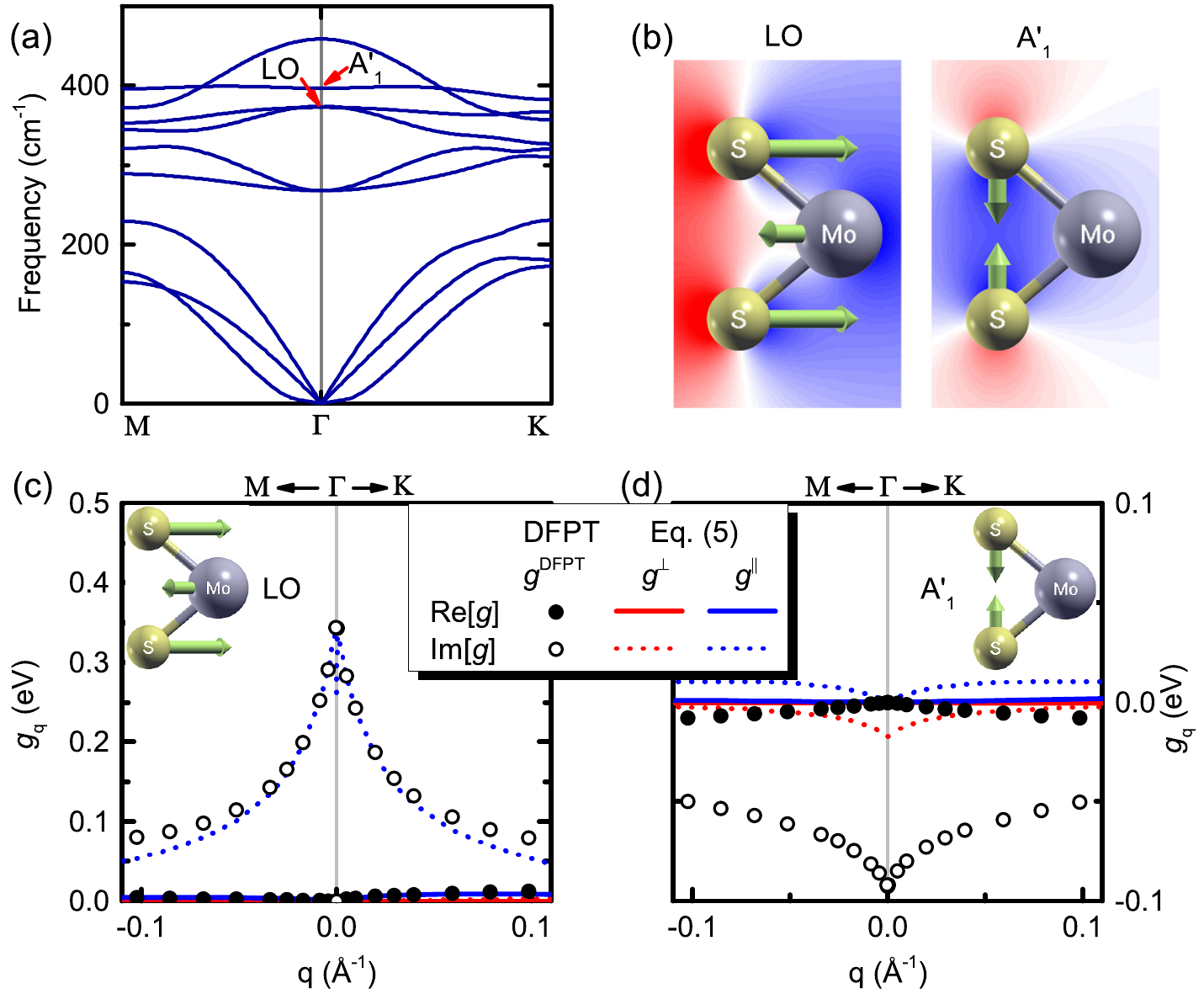}
\caption{\label{fig1} (a) Phonon dispersion of $\text{MoS}_2$, and the vibration patterns of LO and $A_{1}^{'}$ modes. (b) The dipole potential field is illustrated where the red and blue region represents positive and negative potential. $\mathbf{q}$-dependent electron-phonon interaction matrix elements are shown for (c) LO and (d) $A_{1}^{'}$ phonon modes, with electron in the conduction band edge at K point as the initial state. Symbols are DFPT results ($g^{\text{DFPT}}$) while red and blue lines represent the out-of-plane ($g^\perp$) and in-plane ($g^\parallel$) components of Eq. (5), respectively. The difference between DFPT and Eq. (5) result for $A_{1}^{'}$ phonon is attributed to the non-zero short-range contribution.}
\end{figure}

\section{Dipolar electron-phonon interactions in 2D materials}
The 3D \textit{ab initio} Fr{\"o}hlich model\cite{Sjakste2015,Verdi2015} was obtained by solving Poisson equation for a dipole $\mathbf{p}$ in an anisotropic dielectric which leads to a potential field
\begin{equation}
V^{\text{dip,3D}} \left( \mathbf{r} \right) = i\frac{2\pi }{N\Omega}\sum\limits_{\mathbf{q}} \sum\limits_{\mathbf{G\neq q}} \frac{\mathbf{p\cdot (q+G)}e^{i\mathbf{(q+G)\cdot r}}}{\mathbf{(q+G) \cdot \epsilon ^{\infty} \cdot (q+G)}} ,
\end{equation}
where $\epsilon ^{\infty}$ is the high-frequency (electronic) dielectric tensor, and $N\Omega $ is the Born-von Kármán cell size. Here we take $\frac{e}{4\pi\varepsilon_0}$ to be unity, as in atomic units. The $\mathbf{p\cdot(q+G)}$ term suggests that only the longitudinal components of these dipoles contributes to the long-wavelength singularity of Fr{\"o}hlich EPI near $\mathbf{q}=0$. Similarly, in previous 2D Fr{\"o}hlich models, only in-plane dipole components contribute to the non-analytical behavior of EPI due to the $\mathbf{p\cdot q}$ term. However, the out-of-plane vibration of ions also generates a long-range field which is non-analytical and gives EPI a strong $\mathbf{q}$-dependence. This is observed in the small-$\mathbf{q}$ EPI of the homopolar $A_{1}^{'}$ phonon in TMDC, which is the simultaneous out-of-plane vibration of chalcogen atoms in opposite directions, as shown in Figure 1. This long-range field, just like the in-plane term described in 2D Fr{\"o}hlich models, also originates from the Coulomb interaction that should be described in a unified formalism.

To derive the desired formalism, we start by solving the Poisson equation for a point charge $e$ residing in a 2D dielectric. By considering an anisotropic polarizability tensor $\alpha ^{2D}$, the Poisson equation for electrostatic potential $\phi $ becomes \cite{Cudazzo2011}
\begin{equation}
\nabla^2 \phi(\mathbf{r_\parallel ,z}) =  - 4\pi\delta(\mathbf{r}) - 4\pi\delta(z)\mathbf{\nabla _{\parallel } \cdot \alpha ^{\text{2D}} \cdot \nabla _{\parallel }}\phi ( \mathbf{r_\parallel ,z} ),
\end{equation}
where $\bf{r}_\parallel$ and $z$ are in-plane and out-of-plane components of coordinate ${\bf{r}}$. The solution is
\begin{equation}
\phi(\mathbf{r_\parallel ,z}) = \frac{2\pi }{NA}\sum\limits_{\mathbf{q}} \sum\limits_{\mathbf{Q}=\mathbf{q} + \mathbf{G}_\parallel} \frac{e^{-|\mathbf{Q}||z|}e^{i\mathbf{Q \cdot r}}}{|{\mathbf{Q}}| + 2\pi \mathbf{Q \cdot \alpha ^{\text{2D}} \cdot Q}},
\end{equation}
with $A$ and $\mathbf{G}_\parallel$ being the unit cell area and an in-plane reciprocal lattice vector. This is the 2D Fourier transform of Coulomb interaction, with a Keldysh-type dielectric function \cite{Keldysh1979}. The 2D polarizability is computed from macroscopic dielectric tensor \cite{Giannozzi1991} through $\alpha_{{ij}}^{2D} = \frac{c\left( \epsilon_{{ij}}^{\infty} - 1 \right)}{4\pi}$ with $c = \frac{\Omega}{A}$ being the unit cell dimension along perpendicular direction \cite{Berkelbach2013}. This is derived by evaluating the averaged macroscopic dielectric constant of stacked 2D layers, as discussed in Appendix A. The potential field generated by an infinitesimal dipole $\mathbf{p}$ is approximated using the gradient of $\phi(\mathbf{r_\parallel ,z})$ such that ${\phi ^{\text{dip}}} \approx  - \mathbf{p} \cdot \nabla \phi(\mathbf{r_\parallel ,z})$. With this approximation we express the interaction between a dipole $\mathbf{p}$ at origin and an electron as $V^{\text{dip,Q2D}}=-e\phi^{\text{dip}}$ , which is
\begin{equation}
V^{\text{dip,Q2D}}(\mathbf{r_\parallel ,z}) = \frac{2\pi }{NA}\sum\limits_{\mathbf{Q}\equiv\mathbf{q} + \mathbf{G}_\parallel} e^{-|\mathbf{Q}||z|}e^{i\mathbf{Q \cdot r}}\frac{{i{\mathbf{p}} \cdot \left[ {{\mathbf{\hat Q}} + i\mathbf{\hat z}\text{ sgn}(z)} \right]}}{1 + 2\pi \mathbf{\hat Q \cdot \alpha ^{\text{2D}} \cdot \hat Q} |{\mathbf{Q}}|},
\end{equation}
Here ${\mathbf{\hat Q}}$ and ${\mathbf{\hat z}}$ are the unit vectors along $\mathbf{Q}$ and perpendicular directions, respectively, and $\text{sgn}(z)$ is the sign function. The dipole moment of a displaced ion is expressed using Born effective charge as  $e\mathbf{Z}_\kappa^*  \cdot \Delta\mathbf{\tau }_{\kappa \mathbf{R}}$. For a phonon mode $\nu \mathbf{q}$, the displacement of atom $\kappa $ in unit cell $\mathbf{R}$ is $\Delta\mathbf{\tau }_{\kappa \mathbf{R}} = \left( \frac{\hbar }{2N{M_\kappa }\omega_{\nu\mathbf{q}}} \right)^{\frac{1}{2}}e^{i\mathbf{q\cdot R}} \mathbf{e}_{\kappa \nu }(\mathbf{q})$ \cite{Verdi2015,Giustino2017}. Then the 2D dipolar EPI matrix element is
\begin{eqnarray}
g_{mn\nu }^{\text{dip,Q2D}}( \mathbf{k,q} ) &=& i\frac{2\pi }{A}\sum\limits_{\mathbf{Q} = \mathbf{q} + \mathbf{G}_\parallel } \sum\limits_\kappa  {\left(\frac{\hbar}{2N{M_\kappa}\omega_{\nu\mathbf{q}}}\right)}^{\frac{1}{2}} \frac{e^{-i\mathbf{Q \cdot \tau_\kappa}}}{1 + 2\pi \mathbf{\hat Q \cdot \alpha ^{\text{2D}} \cdot \hat Q} |{\mathbf{Q}}|}\nonumber\\
&&\times{\left[ {\mathbf{\hat Q}M_{\kappa mn}(\mathbf{k,Q}) + i{\mathbf{\hat z}}S_{\kappa mn}(\mathbf{k,Q})} \right] \cdot \mathbf{Z}_\kappa^*  \cdot {\mathbf{e}_{\kappa\nu}}(\mathbf{q})},\\
M_{\kappa mn}(\mathbf{k,Q}) &=& \left\langle \psi_{m\mathbf{k+q}}\left| e^{-|\mathbf{Q}||z-z_\kappa|}e^{i\mathbf{Q \cdot r}}\right|\psi _{n\mathbf{k}} \right\rangle,\\
S_{\kappa mn}(\mathbf{k,Q}) &=& \left\langle \psi_{m\mathbf{k+q}}\left| \text{sgn}(z-z_\kappa)e^{-|\mathbf{Q}||z-z_\kappa|}e^{i\mathbf{Q \cdot r}}\right|\psi _{n\mathbf{k}} \right\rangle.
\end{eqnarray}

\begin{figure}
\includegraphics[width=8.6cm]{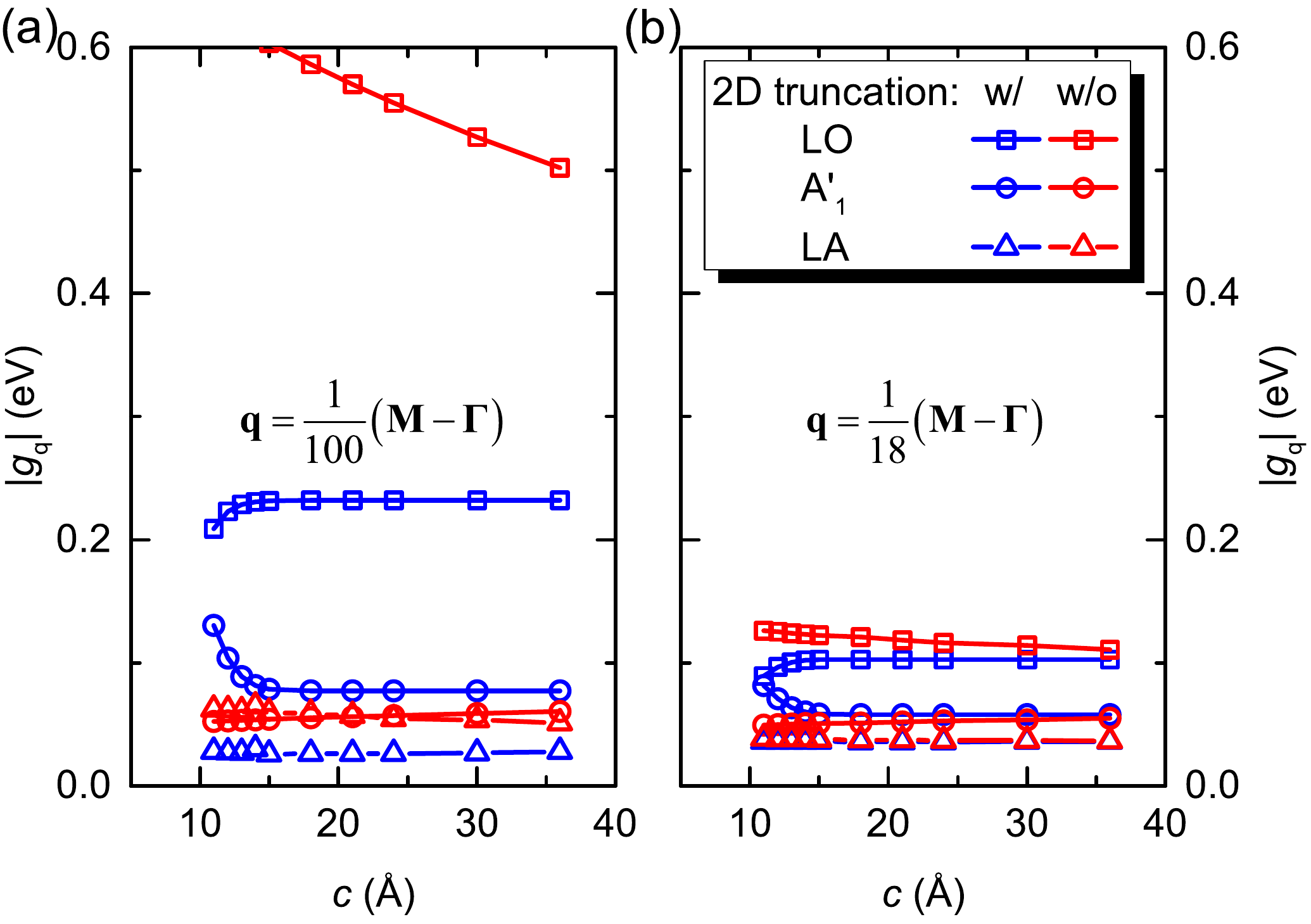}
\label{fig2}
\caption{The long-wavelength EPI for $\text{MoS}_2$ conduction band edge, calculated with different vacuum spacing $c$. EPI is computed for a fixed phonon momentum at (a) 1/100 of and (b) 1/18 of the length between M and $\Gamma$ point in the Brillouin zone. Square, circle and triangle represents EPI with longitudinal optical (LO), homopolar ($A_{1}^{'}$) and longitudinal acoustic (LA) phonon branches. Calculation with and without 2D Coulomb truncation are marked blue and red, respectively.}
\end{figure}
The main differences between this result and previous works are (i) the $i{\mathbf{\hat z}}S_{\kappa mn}(\mathbf{k,Q})$ term describing the out-of-plane dipoles, and (ii) the $e^{-|\mathbf{Q}||z-z_\kappa|}$ factor which couples with the actual wave function span in perpendicular direction. These factors emphasize the quasi-two-dimensional (Q2D) nature of 2D materials, which are 2D-like objects in a 3D world. Since for small $\mathbf{q}$ the potential decays very slowly in perpendicular direction due to the $e^{-|\mathbf{Q}||z-z_\kappa|}$ term, 2D Coulomb truncation is necessary for avoiding spurious coupling between repeating image layers \cite{Huser2013,Qiu2016,Sohier2017}, which is realized by adding a rectangular factor $\theta(l-|z-z_\kappa|)$ with $l=c/2$. In fact, without 2D Coulomb truncation, the EPI is enlarged to more than twice the original strength. Here we show in Figure 2 that due to the $e^{- \left| \mathbf{q} \right|\left| z - z_{\kappa} \right|}$ factor in long-wavelength dipolar EPI, the spurious interaction between electrons in the original layer and the ions in the neighboring image layer decreases slowly for small $\mathbf{q}$. Particularly, this effect will be negligible only if the vacuum separation is much greater than $\frac{1}{\left| \mathbf{q} \right|}$. This means without truncating the Coulomb interaction\cite{Sohier2017}, small $\mathbf{q}$ EPI will always suffer from such spurious interaction, leading to extremely slow convergence with respect to the vacuum size. This effect has also been observed in electron-electron interaction before\cite{Huser2013,Qiu2016}. This is also verified by our DFPT calculation for $\text{MoS}_2$. As shown in Figure 2(a), for both LO, $A_{1}^{'}$ and longitudinal acoustic (LA) phonons, the EPI at $\mathbf{q =}\frac{1}{100}\left( \mathbf{M} - \mathbf{\Gamma} \right)$ without Coulomb truncation is far from convergence as vacuum size $c$ increases, and the relative error can be as large as 200\%. For a slightly larger $\mathbf{q =}\frac{1}{18}\left( \mathbf{M} - \mathbf{\Gamma} \right)$, the deviation is smaller since $e^{- \left| \mathbf{q} \right|\left| z - z_{\kappa} \right|}$ decays faster [See Figure 2(b)]. Still, it requires a much larger vacuum size to reach convergence as compared to the calculation with 2D Coulomb truncation. Since small momentum scattering can be important in polar materials, absence of Coulomb truncation may add spurious error that cannot be easily remedied by increasing vacuum size (See Figure 2) or using denser Brillouin zone sampling.

Here we compared our model dipolar contribution using Eq. (5) with DFPT result $g^{\text{DFPT}}$ computed for $\text{MoS}_2$ conduction band edge. For the real and imaginary part of $g$ to be meaningful, we fixed the gauge by requiring both $\left\langle \psi _{m\mathbf{k+q}}\left| e^{i\mathbf{q \cdot r}} \right|\psi _{n\mathbf{k}} \right\rangle$ and sulfur atom’s longitudinal eigenmode $\mathbf{q \cdot e}_{\kappa\nu}(\mathbf{q})$ to be real and non-negative. Only LO and homopolar $A_{1}^{'}$ phonons have non-vanishing intra-band EPI at $\mathbf{q \to 0}$ \cite{Sohier2016}. The in-plane longitudinal component $g^\parallel$ from $M_{\kappa mn}(\mathbf{k,Q})$, and out-of-plane contribution $g^\perp$ from $S_{\kappa mn}(\mathbf{k,Q})$, are plotted separately. The LO EPI approaches a constant value of 0.34 eV while DFPT result is zero at $\mathbf{q=0}$ due to periodic boundary condition. As shown in Figure 1 (c) for LO phonon, the in-plane $g^\parallel$ correctly reproduces the long-wavelength behavior, while $g^\perp$ has negligible contribution due to vanishing out-of-plane vibration and smaller effective charge components ($Z_{\text{S},\perp}^*=0.034$ and $Z_{\text{Mo},\perp}^*=0.068$, versus $Z_{\text{S},\parallel}^*=0.494$ and $Z_{\text{Mo},\parallel}^*=0.989$). For the $A_{1}^{'}$ phonon branch in Figure 1(d), dipolar EPI also contribute to the cusp singularity near $\mathbf{q \to 0}$. Although the in-plane component vanish at $\mathbf{q \to 0}$, both in-plane and out-of-plane vibrations add a cusp to $A_{1}^{'}$ EPI with comparable slope because $Z_{\text{S},\parallel}^* \gg Z_{\text{S},\perp}^*$. For systems with significant out-of-plane Born effective charge such as graphene derivatives, ${g^ \perp }$ could dominate over ${g^ \parallel }$ in homopolar phonons, as dicussed in Appendix B. Due to such cusp singularity, $A_{1}^{'}$ EPI strength reduces significantly as $|\mathbf{q}|$ increases. This effect was not addressed in previous models and suggests that a constant approximation \cite{Schmid1974,Fivaz1967} or Fourier interpolation could fail for $A_{1}^{'}$ phonon without dipolar correction. We noticed that a recent work \cite{Singh2020} suggested a dipolar EPI from the highest phonon branch of $\text{MoS}_2$. However, it vanishes in our model due to mirror symmetry of $\text{MoS}_2$ monolayer, as verified in DFPT calculation. Such coupling may be important in stacked layers and can be described within our formalism. The presence of ${g^\perp}$ contribution is the consequence of 2D materials’ Q2D nature, where vibrations in the third dimension play important role in electron-phonon coupling.

\begin{figure}
\includegraphics[width=8.6cm]{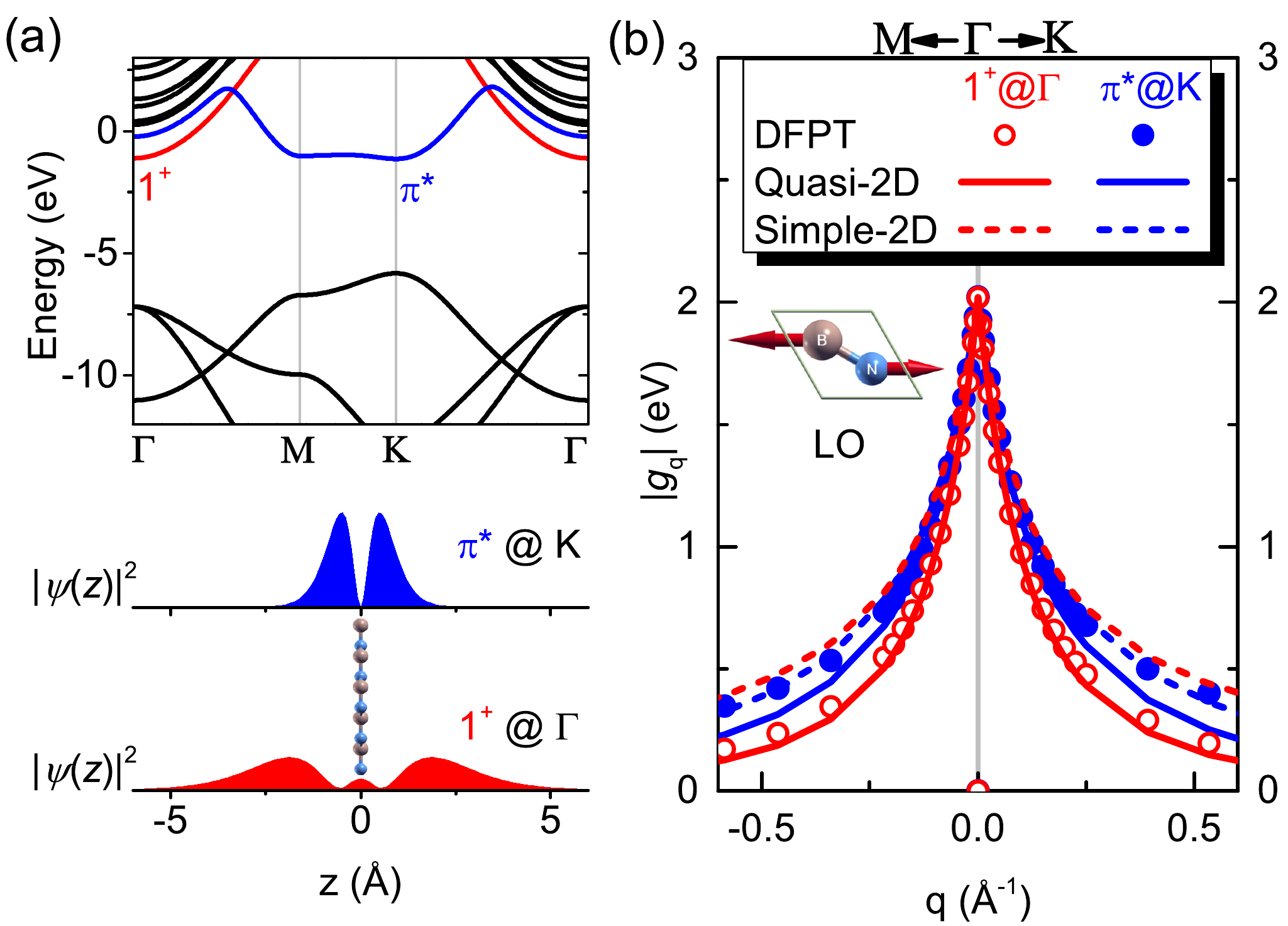}
\caption{\label{fig3} (a) Electron band structure of hBN, with vacuum level at 0 eV. The red and blue lines represent the first image potential state ($1^+$), and antibonding $\pi^*$ state consisting of $p_z$ orbitals, respectively. The plane-averaged wave functions are illustrated along out-of-plane z direction, which shows that $1^+$ state has a wider distribution. (b) $\mathbf{q}$-dependent electron-LO interaction strength, where $1^+$ and $\pi^*$ states are marked red and blue, respectively. Symbols, solid and dashed lines represent the DFPT result, quasi-2D model given by Eq. (5), and simple-2D model without the $e^{-|\mathbf{Q}||z-z_\kappa|}$ factor.}
\end{figure}

In addition to the Keldysh-type screening length $\mathbf{r}_{\text{eff}}=2\pi \mathbf{\hat Q \cdot \alpha^{\text{2D}} \cdot \hat Q}$, the $e^{-|\mathbf{Q}||z-z_\kappa|}$ factor induces another length scale contributing to the singularity near $\mathbf{q \to 0}$. This effect has been discussed in electron-electron interaction model \cite{Deng2015}, while its role in EPI was only considered empirically \cite{Kaasbjerg2012}. This length scale is related to the effective thickness of electronic state, and becomes non-negligible when comparable with $\mathbf{r}_{\text{eff}}$. To illustrate this, we computed the LO EPI for 2D hexagonal boron nitride (hBN) whose conduction band at $\Gamma$ point is an image potential state ($1^+$) like that in graphene, while the conduction band at K point is simply a $\pi^*$ state consisting of ${p_z}$ orbitals, as shown in Figure 3(a). While the $\pi^*$ state is tightly bound to the atoms, the $1^+$ state has a wider out-of-plane distribution where the $e^{-|\mathbf{Q}||z-z_\kappa|}$ factor shows different impact. As shown in Figure 3(b), the LO EPI with $1^+$ state weakens faster than that for $\pi^*$ state as $|\mathbf{q}|$ increases. While our model $g^{\text{dip,Q2D}}$ correctly captures the different decay rate, simple 2D model without $e^{-|\mathbf{Q}||z-z_\kappa|}$ shows very small difference and overestimates coupling between LO phonon and $1^+$ state. Our results demonstrate that the Q2D nature of 2D materials should be considered when studying 2D materials with significant electronic thickness.

The dipolar contribution described here can be subtracted from the DFPT electron-phonon interaction potential and enables Fourier interpolation \cite{Ponce2015,Brunin2020,Brunin2020a}. Alternatively, $g^{\text{dip,Q2D}}$ could be subtracted from EPI matrix elements after simplification and enables Wannier-Fourier interpolation to reduce computational cost \cite{Verdi2015,Zhou2020,Jhalani2020}. The simplified $g^{\text{dip,Q2D}}$ must satisfy the following criteria: (i) all quantities to be Wannier-interpolated are smooth in $\mathbf{k}$; (ii) it reduces to Eq. (5) in long-wavelength limit; (iii) it satisfies the Hermitian relation $g_{nm\nu }^{\text{dip,Q2D}}(\mathbf{k+q,-q})=\left[ {g_{mn\nu }^{\text{dip,Q2D}}(\mathbf{k,q})} \right]^*$. Assisted by smoothness of Wannier-gauge \cite{Wang2006} we are able to identify such approximation, which is detailed in Appendix C. We note here that the resulting approximation only needs Wannier interpolation in electron Brillouin zone, and the overlap matrix $M_{\kappa mn}(\mathbf{k,q})$ is identical to that in the 3D \textit{ab initio} Fr{\"o}hlich model \cite{Verdi2015}. After subtracting $g^{\text{dip,Q2D}}$ from the electron-phonon matrix $g$, the remainder $g^{\text{rem,Q2D}}=g-g^{\text{dip,Q2D}}$ becomes smoother and Wannier-Fourier interpolation can be applied.

\begin{figure}
\includegraphics[width=8.6cm]{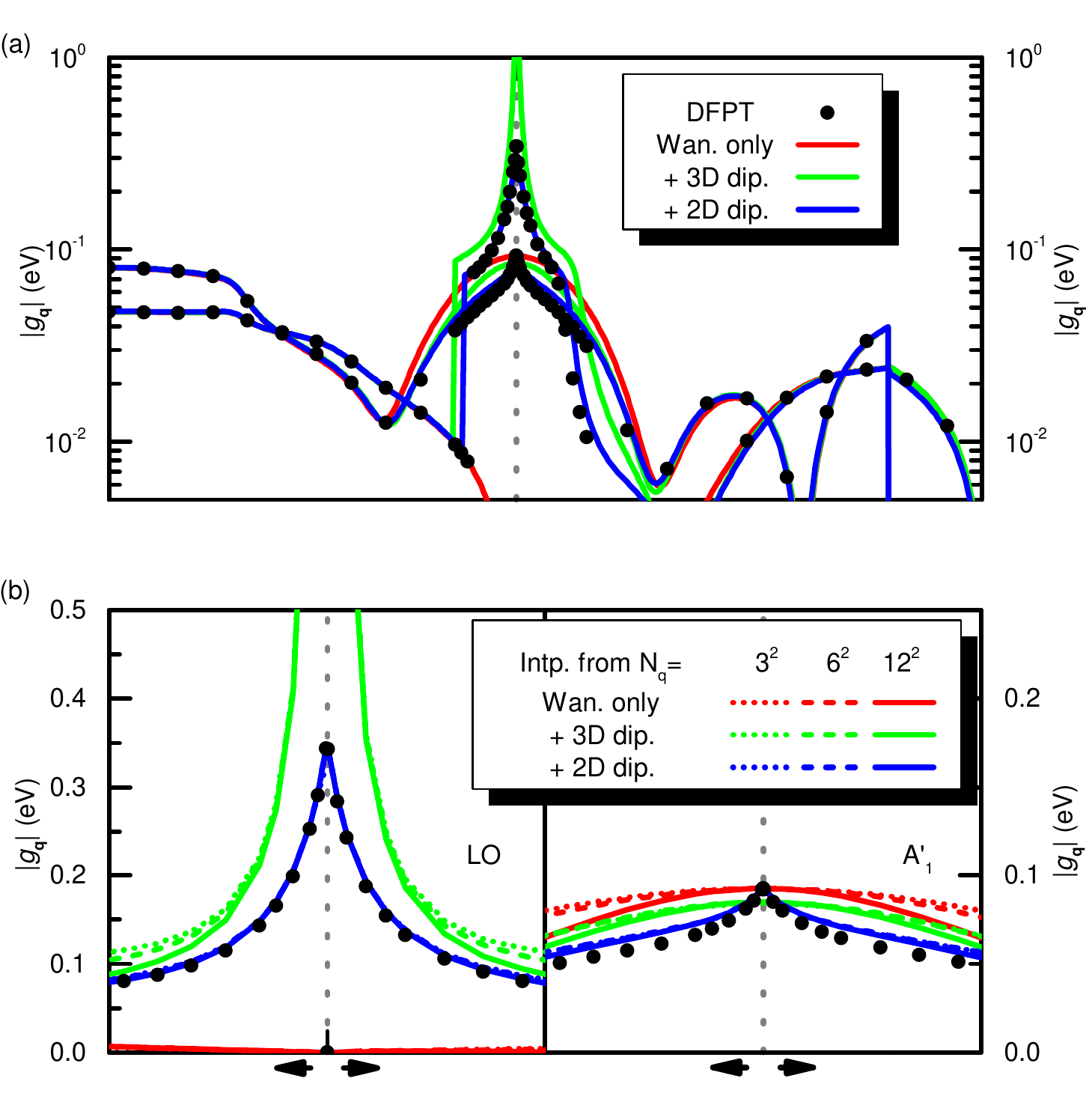}
\caption{\label{fig4} Electron phonon matrix elements calculated from DFPT (symbols) and Wannier-Fourier interpolation schemes without dipolar correction (Wan. only, red lines), with 3D Fr{\"o}hlich correction (+3D dip., green lines), and with 2D dipolar correction in this work (+2D dip., blue lines). The matrix elements are computed for $\text{MoS}_2$ with conduction band edge as initial state, and for phonon momentum $\mathbf{q}$ (a) along M-$\Gamma$-K high symmetry path and (b) in the vicinity of $\Gamma$ point.}
\end{figure}

With our implementation in {\sc Quantum ESPRESSO} \cite{Giannozzi2009,Giannozzi2017} and EPW \cite{Ponce2016}, we compute the EPI for $\text{MoS}_2$ for demonstration. As shown in Figure 4 the Wannier-Fourier interpolation without correction leads to a vanishing LO EPI, because DFPT result at $\Gamma$ point in the absence of dipolar correction is zero due to periodicity. It cannot reproduce the cusp singularity for $A_{1}^{'}$ phonon either, and artificially smoothens $g$. The 3D Fr{\"o}hlich correction produces a diverging LO EPI near $\mathbf{q \to 0}$ which reduces slowly with increased initial $\mathbf{q}$-grid size. The model presented in this work reproduces the small-$\mathbf{q}$ DFPT results even with a $3\times 3\times 1$ initial $\mathbf{q}$-grid. This demonstrated the intrinsic difference between EPIs in 2D and 3D materials.

\section{Phonon-limited intrinsic carrier mobilities}
\begin{figure}
\includegraphics[width=8.6cm]{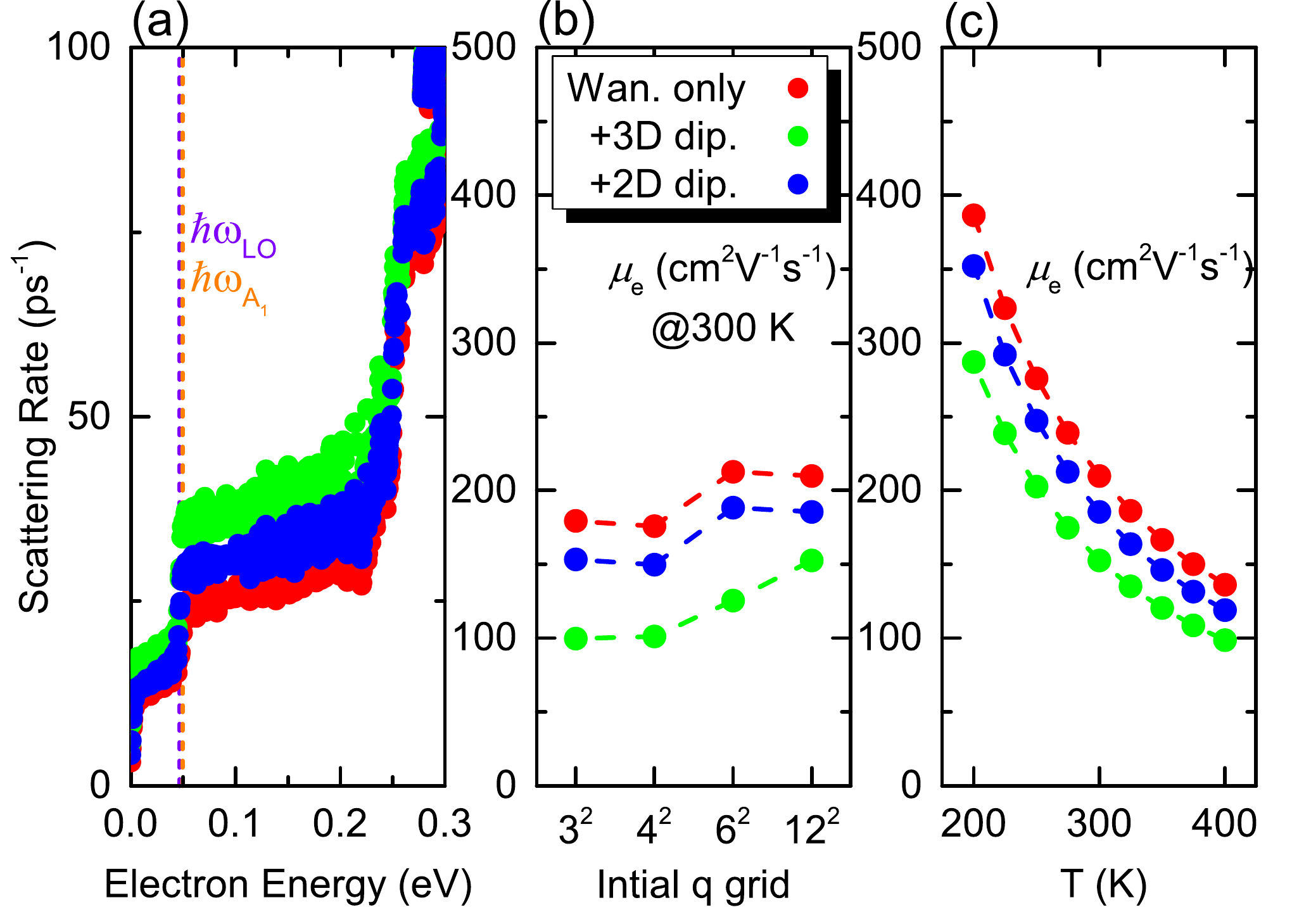}
\label{fig5}
\caption{The ERTA results for electrons in $\text{MoS}_2$ including (a) scattering rate, (b) convergence with respect to initial $\mathbf{q}$-grid, and (c) temperature dependence of ERTA mobility.}
\end{figure}
In addition to these $\mathbf{k(q)}$-dependent quantities, it is also important to see how dipolar corrections affect quantities that require integration over the whole Brillouin zone. To this end, we computed the $\text{MoS}_2$ intrinsic electron mobility by iteratively solving Boltzmann transport equation (BTE) \cite{Ponce2018}. All the DFT and DFPT calculations were performed using PBE functional \cite{Perdew1996} and norm-conserving pseudopotentials from PseudoDojo \cite{VanSetten2018} with a kinetic energy cutoff of 70 Ry. Ground state charge density is obtained using $12\times 12\times 1$ $\mathbf{k}$-grid, and the same $\mathbf{k}$-grid is used for Wannier function construction. This grid is sufficient to reach convergence using 2D dipolar correction, as shown in Figure 5. EPI and band structures were interpolated to a uniform $360\times 360\times 1$ $\mathbf{k(q)}$-grid where convergence was achieved. Both energy relaxation time approximation (ERTA) and self-consistent iterative solution of BTE (IBTE) were used to compute the mobilities in Figure 6. As shown in Figure 5(a) and 6(a), the scattering is underestimated without dipolar correction, leading to a mobility overestimation similar to 3D polar materials \cite{Zhou2016}. Meanwhile, 3D dipolar correction overestimates the scattering and reduces mobility. These effects become more pronounced for electron energy higher than the LO and $A_{1}^{'}$ phonon energies after onset of phonon emission processes, as shown in Figure 5(a). We also found that the ERTA electron mobility at 300 K is changed by only 1\% when increasing the initial $\mathbf{q}$-grid from $6\times 6\times 1$ to $12\times 12\times 1$. The converged electron mobility at 300 K using 2D correction was 176.6 $\text{cm}^{2} \text{V}^{-1} \text{s}^{-1}$, as compared to 189.4 $\text{cm}^{2} \text{V}^{-1} \text{s}^{-1}$ without correction and 158.0 $\text{cm}^{2} \text{V}^{-1} \text{s}^{-1}$ with 3D corretion. We also computed the mobility using 2D Fr{\"o}hlich model proposed by Sohier et al \cite{Sohier2016}, which is 174.3 $\text{cm}^{2} \text{V}^{-1} \text{s}^{-1}$ and is very close to our model. This is because Sohier's model correctly captures the in-plane component of dipolar interaction $g^{\parallel}$ which has the dominant contribution due to large in-plane Born effective charges. The out-of-plane term which is missing in their 2D Fr{\"o}hlich model can have more significant impact in systems with stronger out-of-plane dipolar EPI, where accounting its effect will be important to understand the electron-phonon interaction.  Since the acoustic phonon and inter-valley scatterings dominate in $\text{MoS}_2$ \cite{Sohier2018}, the dipolar correction has limited impact in this case. However, it is expected to qualitatively alter the prediction for materials with stronger dipolar EPI, such as TMDCs with large Born effective charges \cite{Cheng2018}. We noticed that additional $\mathbf{q}$ sampling were employed to improve the prediction of 3D dipolar correction for InSe \cite{Li2019,Ma2020}, and we have also tested its performance for $\text{MoS}_2$, as detailed in Appendix D.

\begin{figure}
\includegraphics[width=8.6cm]{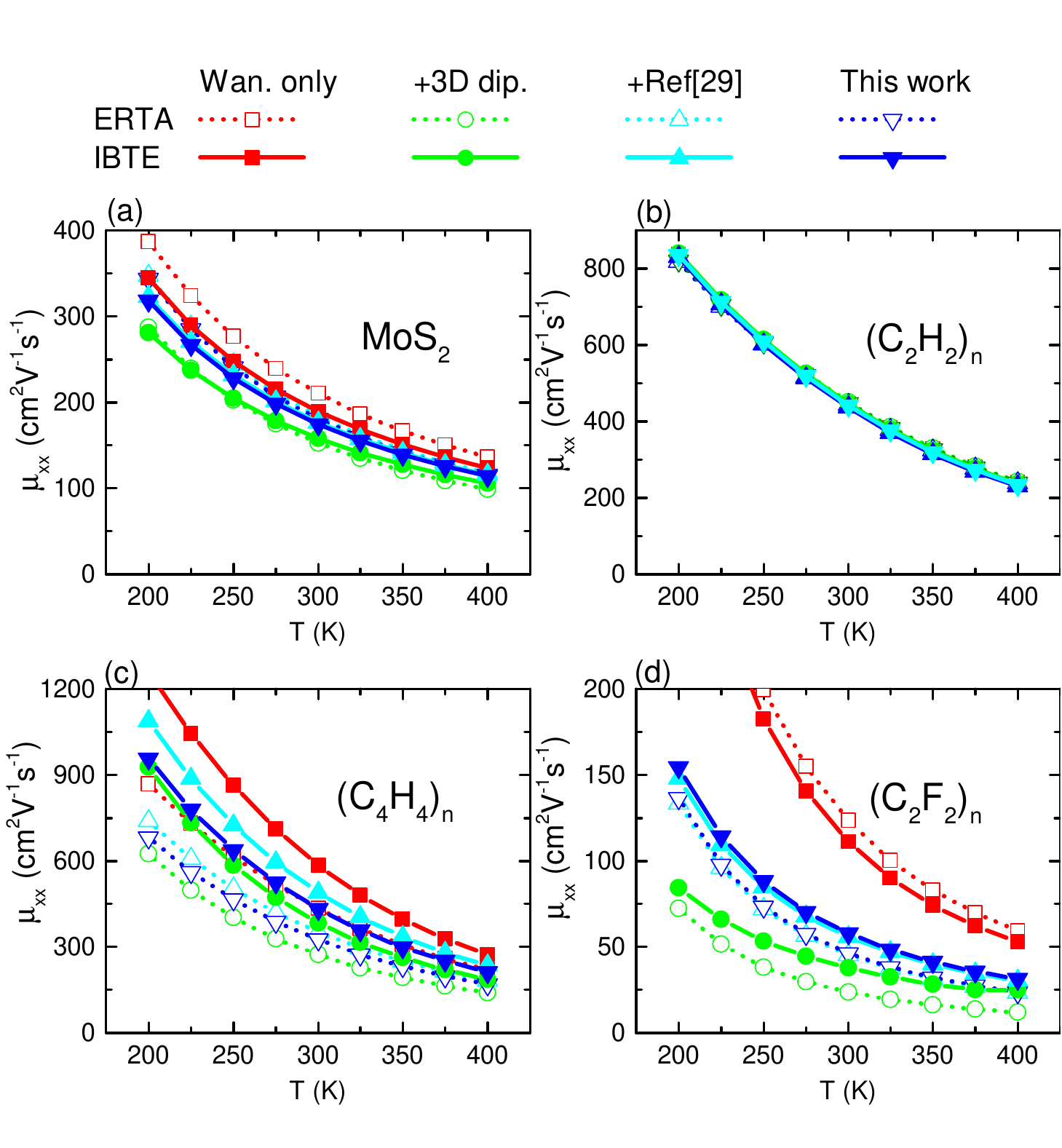}
\caption{\label{fig6} Phonon-limited mobilities of electrons in (a) $\text{MoS}_2$, and holes in (b) chair-like graphane [$(\text{C}_2\text{H}_2)_n$], (c) boat-like graphane [$(\text{C}_4\text{H}_4)_n$], and (d) chair-like fluorinated graphene [$(\text{C}_2\text{F}_2)_n$]. Both results from energy relaxation time approximation (ERTA) and iterative solution of Boltzmann transport equation (IBTE) are shown. Electron-phonon matrix elements are interpolated using different Wannier-Fourier interpolation schemes, including the original Wannier-Fourier interpolation schemes without dipolar correction (Wan. only, red symbols), with 3D Fr{\"o}hlich correction (+3D dip., green symbols), with 2D Fr{\"o}hlich model by Sohier et al \cite{Sohier2016} (+Ref[29], cyan symbols) and 2D dipolar correction in this work (+2D dip., blue symbols).}
\end{figure}

More interestingly, we found that dipolar interaction can be important in some graphene derivatives. Specifically, we computed the hole mobilities of chair-like [$(\text{C}_2\text{H}_2)_n$], boat-like [$(\text{C}_4\text{H}_4)_n$] graphane \cite{Sofo2007} and chair-like fluorinated graphene [$(\text{C}_2\text{F}_2)_n$]. A kinetic energy cutoff of 90 Ry was used. EPI and band structures were interpolated from $12\times 12\times 1$ to $360\times 360\times 1$ $\mathbf{k}$-($\mathbf{q}$-)grid for $(\text{C}_2\text{H}_2)_n$ and $(\text{C}_2\text{F}_2)_n$, and from $7\times 14\times 1$ to $200\times 340\times 1$ for $(\text{C}_4\text{H}_4)_n$, respectively. Contributions from low frequency ($\leq 10 \text{ cm}^{-1}$) phonons were removed to avoid numerical instability. Although the covalent C-H bond is usually assumed nonpolar, the slight electronegativity difference between C and H still allows non-zero Born effective charges. And in the presence of non-zero Born effective charges, dipolar EPI can play an important role just like in polar insulators. In $(\text{C}_2\text{H}_2)_n$, the in-plane $Z_\parallel^*$ is only 0.014 while $Z_\perp^*$ is 0.102. The out-of-plane optical phonon involves high frequency C-H bond stretching and is hardly activated. Therefore, the overall dipolar contribution to scattering is weak, and the mobility predicted from different interpolation schemes are similar. In $(\text{C}_4\text{H}_4)_n$ however, the H atom has an off-diagonal charge $Z_{{\text{H}},xz}^*=0.120$ due to the reduced symmetry, which means the low-frequency flexural optical phonon generates in-plane dipoles with strong dipolar EPI. Consequently, hole mobility at 300 K is overestimated by 20\% without dipolar correction, while the 3D correction underestimates it by 22\%. For $(\text{C}_2\text{F}_2)_n$, Born effective charges are even larger ($Z_\parallel^*=0.290$, $Z_\perp^*=0.330$), so the dipolar EPI is more significant and hole mobility at 300 K is overestimated by 99\% in the absence of dipolar correction, while 3D dipolar correction underestimate it by 33\%. Such observation in graphene derivatives suggests that even in carbon-based systems, dipolar EPI may play major role and correct treatment is necessary. While dipolar effect can be weak in cases like $(\text{C}_2\text{H}_2)_n$, certain carbon-based materials can exhibit strong dipolar EPI just like polar insulators and it is important to correctly consider such effect in calculations. Without including dipolar effects, the calculations may result in inaccurate property predictions. Similar phenomena may be observed in other carbon-based or organic 2D materials, such as covalent organic frameworks \cite{Feng2012}.

\section{Conclusion}
In conclusion, we proposed a unified \textit{ab initio} formalism describing dipolar electron-phonon interaction in 2D materials by incorporating both longitudinal and out-of-plane dipoles. Our observation demonstrated the importance of their quasi-2D nature in understanding 2D materials. The proposed formalism improves the accuracy and efficiency of \textit{ab initio} electron-phonon interaction calculations through Wannier-Fourier interpolation. We demonstrated the implementation and application by computing the intrinsic electron mobility of $\text{MoS}_2$ and hole mobility of graphene derivatives. We found dipolar interaction is important not only in typical polar materials, but also in certain graphene derivatives. This method can be useful for other relevant studies such as optical properties and polaron formation.

\begin{acknowledgments}
This work is supported by Agency for Science, Technology and Research (A*STAR) of Singapore (1527200024). Computational resources are provided by A*STAR Computational Resource Centre (A*CRC).
\end{acknowledgments}

\appendix
\section{Coulomb interaction in quasi-2D system}

Here we solve the Poisson equation Eq. (2). To this end, we first perform Fourier transformation in all three directions with $\mathbf{q}$ being the in-plane momentum and $\mathbf{G}_{\mathbf{\bot}}$ being the out-of-plane momentum. In
periodic boundary condition this is defined as by $\phi\left( \mathbf{Q,}\mathbf{G}_{\bot} \right) = \int{\phi\left( \mathbf{r}_{\parallel},z \right)\exp\left( - i\mathbf{Q \cdot}\mathbf{r}_{\parallel} - iG_{\bot}z \right)d\mathbf{r}_{\parallel}dz}$ whose inverse transformation is $\phi\left( \mathbf{r}_{\parallel},z \right) = \frac{1}{N A c}\sum_{\mathbf{Q,}\mathbf{G}_{\bot}}{\phi\left( \mathbf{Q,}\mathbf{G}_{\perp} \right)\exp\left( i\mathbf{Q \cdot}\mathbf{r}_{\parallel} + iG_{\perp}z \right)}$. Then Eq. (2) becomes
\begin{equation}
- \left( \mathbf{Q}^{\mathbf{2}}\mathbf{+}\mathbf{G}_{\mathbf{\perp}}^{\mathbf{2}} \right)\phi\left( \mathbf{Q},\mathbf{G}_{\mathbf{\perp}} \right) = - 4\pi + 4\pi\ \mathbf{Q} \cdot \mathbf{\alpha}^{2D} \cdot \mathbf{Q}\frac{1}{c}\sum_{\mathbf{G}_{\perp}}{\phi\left( \mathbf{Q,}\mathbf{G}_{\perp} \right)\exp\left( iG_{\perp}z \right)}.
\end{equation}
This leads to
\begin{eqnarray}
\sum_{\mathbf{G}_{\perp}}{\phi\left( \mathbf{Q,}\mathbf{G}_{\perp} \right)\exp\left( iG_{\perp}z \right)}&=&\sum_{\mathbf{G}_{\perp}}\frac{4\pi}{\mathbf{Q}^{\mathbf{2}}\mathbf{+}\mathbf{G}_{\mathbf{\perp}}^{\mathbf{2}}}\nonumber\\
&&-\mathbf{Q} \cdot \mathbf{\alpha}^{2D} \cdot \mathbf{Q}\sum_{\mathbf{G}_{\perp}}\frac{4\pi}{\mathbf{Q}^{\mathbf{2}}\mathbf{+}\mathbf{G}_{\mathbf{\perp}}^{\mathbf{2}}}\frac{1}{c}\sum_{\mathbf{G}_{\perp}}{\phi\left( \mathbf{Q,}\mathbf{G}_{\perp} \right)\exp\left( iG_{\perp}z \right)},
\end{eqnarray}
whose solution is
\begin{equation}
\frac{1}{c}\sum_{\mathbf{G}_{\perp}}{\phi\left( \mathbf{Q,}\mathbf{G}_{\perp} \right)\exp\left( iG_{\perp}z \right)}\mathbf{=}\frac{2\pi}{\left| \mathbf{Q} \right| + 2\pi\mathbf{Q} \cdot \mathbf{\alpha}^{2D} \cdot \mathbf{Q}}.
\end{equation}
Substituting Eq. (A3) into (A1) and performing inverse Fourier transformation
\begin{eqnarray}
\phi\left( \mathbf{r}_{\parallel},z \right) &=& \frac{1}{{NAc}}\sum_{\mathbf{Q,}\mathbf{G}_{\perp}}{\phi\left( \mathbf{Q,}\mathbf{G}_{\perp} \right)\exp\left( i\mathbf{Q \cdot}\mathbf{r}_{\parallel} + iG_{\perp}z \right)} \nonumber\\
&=& \frac{1}{{NAc}}\sum_{\mathbf{Q,}\mathbf{G}_{\perp}}{\frac{4\pi\exp\left( iG_{\perp}z \right)}{\mathbf{Q}^{\mathbf{2}}\mathbf{+}\mathbf{G}_{\mathbf{\perp}}^{\mathbf{2}}}\frac{\exp\left( i\mathbf{Q \cdot}\mathbf{r}_{\parallel} \right)}{\left| \mathbf{Q} \right| + 2\pi\mathbf{Q} \cdot \mathbf{\alpha}^{2D} \cdot \mathbf{Q}}} \nonumber\\
&=& \frac{2\pi}{NA}\sum_{\mathbf{Q}}\frac{e^{- \left| \mathbf{Q} \right|\left| z \right|}e^{i\mathbf{Q} \cdot \mathbf{r}}}{\left| \mathbf{Q} \right| + 2\pi\mathbf{Q} \cdot \mathbf{\alpha}^{2D} \cdot \mathbf{Q}},
\end{eqnarray}
which is the Eq. (3) in the main text.

To compute the polarizability from first principles, we consider a two-dimensional insulator in a periodic cell with vacuum separating each layer. Assuming the electronic polarization is localized in the $z = 0$ plane and only polarize along the plane such that when there is a homogeneous electric field $\mathbf{E}$, the electronic polarization is determined by a 2-by-2 2D polarizability tensor $\mathbf{\alpha}^{2D}$ via
\begin{equation}
\mathbf{P}_{\mathbf{e}}\left( \mathbf{r} \right) = \delta\left( z \right)\mathbf{\alpha}^{2D}\cdot \mathbf{E}.
\end{equation}
When all 2D layers are separated by $c$ from each other, the total macroscopic polarization can be computed as an average in the unit cell
\begin{equation}
\mathbf{P}^{\text{mac}}\mathbf{=}\frac{1}{\Omega}\int_{\text{uc}}{\mathbf{P}_{\mathbf{e}}\left( \mathbf{r} \right)d\mathbf{r}}\mathbf{=}\frac{1}{c}\mathbf{\alpha}^{2D}\mathbf{\cdot}\mathbf{E}.
\end{equation}
Since the macroscopic dielectric tensor is defined as $\mathbf{E} =\mathbf{\epsilon}^{\mathbf{\infty}}\mathbf{\cdot}\mathbf{E}\mathbf{-}4\pi\mathbf{P}^{\text{mac}}$ \cite{Giannozzi1991}, we could connect polarizability $\mathbf{\alpha}^{2D}$ with macroscopic dielectric tensor $\mathbf{\epsilon}^{\mathbf{\infty}}$ through
\begin{equation}
\alpha_{{ij}}^{2D} = \frac{c\left( \epsilon_{{ij}}^{\infty} - 1 \right)}{4\pi}.
\end{equation}

Like its dipolar counterpart, the quadrupolar contribution could also be derived from the electrostatic potential from Eq. (3) in the main text. This is achieved through
\begin{eqnarray}
&&V^{\text{quad,Q2D}}\left( \mathbf{r}_{\mathbf{\parallel}},z \right) = \frac{2\pi}{{NA}}\sum_{\mathbf{q}}{\sum_{\mathbf{Q} = \mathbf{q} + \mathbf{G}_{\mathbf{\parallel}}}{\mathbf{\nabla} \cdot \mathcal{Q} \cdot \mathbf{\nabla}\frac{e^{- \left| \mathbf{Q} \right|\left| z \right|}e^{i\mathbf{Q} \cdot \mathbf{r}}}{\left| \mathbf{Q} \right| + 2\pi\mathbf{Q} \cdot \mathbf{\alpha}^{2D} \cdot \mathbf{Q}}}}\nonumber\\
&=& \frac{2\pi}{{NA}}\sum_{\mathbf{q}}\sum_{\mathbf{G}_{\mathbf{\parallel}}}\left\{ \left\lbrack i\hat{\mathbf{Q}}\mathbf{- \ }\hat{\mathbf{z}}\text{sgn}\left( z \right) \right\rbrack \cdot \mathcal{Q}\mathbf{\cdot}\left\lbrack i\hat{\mathbf{Q}}\mathbf{- \ }\hat{\mathbf{z}}\text{sgn}\left( z \right) \right\rbrack\mathbf{+}2\hat{\mathbf{z}}\mathcal{\cdot Q}\mathbf{\cdot \ }\hat{\mathbf{z}}\delta\left( z \right) \right\}\nonumber\\
&&\times \frac{\left| \mathbf{Q} \right|^{2}e^{- \left| \mathbf{Q} \right|\left| z \right|}e^{i\mathbf{Q} \cdot \mathbf{r}}}{\left| \mathbf{Q} \right| + 2\pi\mathbf{Q} \cdot \mathbf{\alpha}^{2D} \cdot \mathbf{Q}},
\end{eqnarray}
with $\mathcal{Q}$ being a tensor representing the quadrupole. This term has a $\mathcal{O}\left( \left| \mathbf{q} \right| \right)$ contribution to the EPI. Then the quadrupolar EPI matrix element can be obtained similar to the dipolar term, or directly evaluated\cite{Brunin2020a,Brunin2020,Jhalani2020,Park2020}.

In addition to the quadrupolar term, the Born effective charge also has a $\mathcal{O}\left( \left| \mathbf{q} \right| \right)$ contribution due to the self-consistent change in Hartree-exchange-correlation potential\cite{Brunin2020a,Brunin2020}. We conjecture that this term can be similarly obtained for 2D materials, by multiplying the Born effective charge in 3D model with the dimensionality factor, which is $2\pi c\left| \mathbf{Q} \right|e^{- \left| \mathbf{Q} \right|\left| z \right|}$.

\section{Out-of-plane dipolar EPI in hydrogenated and fluorinated graphene}
\begin{figure}
\includegraphics[width=8.6cm]{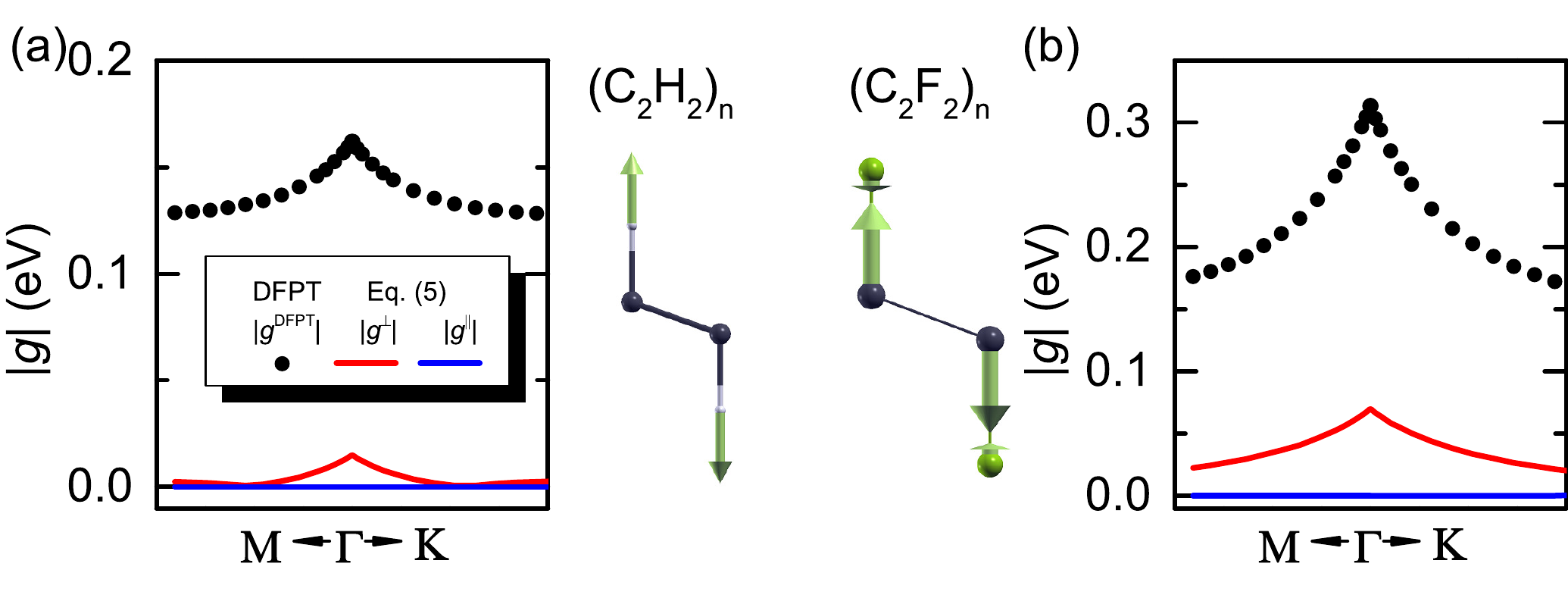}
\label{fig7}
\caption{In (a) chair-like graphane [$(\text{C}_2\text{H}_2)_n$] and (b) fluorinated graphene [$(\text{C}_2\text{F}_2)_n$], the out-of-plane contribution $g^{\perp}$ (red curves) becomes much more significant than in-plane contribution $g^{\parallel}$ (blue curves) for the out-of-plane $A_{1g}$ phonons. This is attributed to the relatively large out-of-plane Born effective charge components in $(\text{C}_2\text{H}_2)_n$ and $(\text{C}_2\text{F}_2)_n$ and vanishing in-plane vibration in $A_{1g}$ modes.}
\end{figure}

Since the out-of-plane Born effective charge components are significant in hydrogenated and fluorinated graphene, we inspect their contribution to the $A_{1g}$ phonons in these systems.  Because the valence band edge studied here is degenerate and choice of initial state becomes arbitrary, we only show the average $|g|$ of these states. As show in Figure 7, the out-of-plane $g^{\perp}$ contribution is much more significant than the in-plane $g^{\parallel}$ contribution. This results from the large out-of-plane Born effective charge [0.102 for $(\text{C}_2\text{H}_2)_n$ and 0.330 for $(\text{C}_2\text{F}_2)_n$], which is much greater than the case of $\text{MoS}_2$ where $Z_{\text{S},\perp}^*=0.034$ and $Z_{\text{Mo},\perp}^*=0.068$.

\section{Implementation into Wannier-Fourier interpolation}
Here we derive the simplified expression for $g^{\text{dip,Q2D}}$ which can be used in Wannier-Fourier interpolation. To satisfy criteria (i) and (ii), we first take the approximation $e^{-|\mathbf{Q}||z-z_\kappa|} \to 1$. After this simplification, we proceed to find the long-wavelength approximation of $M_{\kappa mn}(\mathbf{k,Q})$ and $S_{\kappa mn}(\mathbf{k,Q})$ in Wannier basis defined by Wannier functions $w_{m\mathbf{R}_e}(\mathbf{r})=\frac{1}{N}\sum\limits_{n\mathbf{k}} e^{-i\mathbf{k\cdot R}_e}[ U_\mathbf{k}]_{nm} \psi_{n\mathbf{k}}(\mathbf{r})$. We also define an intermediate, Wannier-derived Bloch-like state $w_{m\mathbf{k}}(\mathbf{r})=\sum\limits_{\mathbf{R}_e} e^{-i{\mathbf{k\cdot R}_e}} w_{m\mathbf{R}_e}(\mathbf{r}) $, which is smooth in $\mathbf{k}$. It is connected to Bloch state through simple rotation $\psi_{n\mathbf{k}}(\mathbf{r})=\sum\limits_m [U_\mathbf{k}^\dag]_{nm}w_{m\mathbf{k}}(\mathbf{r})$. For simplicity we denote the $\mathbf{q=0}$ matrix element of $\text{sgn}(z-z_\kappa)$ in $\psi _{n\mathbf{k}}$ basis as$\left[ {\tilde S}_\kappa (\mathbf{k}) \right]_{mn} \equiv S_{\kappa mn}(\mathbf{k,0})$, and similarly ${\tilde S_\kappa }(\mathbf{R}_e)$ and $\tilde S_\kappa^{(W)}(\mathbf{k})$ for $w_{m\mathbf{R}_e}$ and $w_{m\mathbf{k}}$ bases. They are connected through
\begin{equation}
\tilde S_\kappa(\mathbf{R}_e)=\frac{1}{N}\sum\limits_{\mathbf{k}_0 \in \mathcal{K}} {e^{-i{\mathbf{k}_0} \cdot {{\mathbf{R}}_e}}\tilde S_\kappa ^{(W)}\left( \mathbf{k} \right)} =\frac{1}{N}\sum\limits_{\mathbf{k}_0 \in \mathcal{K}} e^{-i\mathbf{k}_0 \cdot \mathbf{R}_e}\left[ U_{\mathbf{k}_0}^\dag S_\kappa (\mathbf{k}_0)U_{\mathbf{k}_0} \right]_{mn}.
\end{equation}
Here the Wannier-basis matrix elements are obtained using $S_\kappa (\mathbf{k}_0)$ on a coarse grid $\mathbf{k}_0 \in \mathcal{K}$. $\tilde S_\kappa (\mathbf{k})$ for any $\mathbf{k}$ is then obtained through Wannier-interpolation
\begin{equation}
\tilde S_\kappa (\mathbf{k})=U_\mathbf{k}\tilde S_\kappa ^{(W)}(\mathbf{k})U_\mathbf{k}^\dag =U_\mathbf{k}\sum\limits_{\mathbf{R}_e} {e^{i\mathbf{k} \cdot \mathbf{R}_e}\tilde S_\kappa (\mathbf{R}_e)} U_\mathbf{k}^\dag .
\end{equation}
If we simply approximate $S_{\kappa mn}(\mathbf{k,Q})$ with $\tilde S_\kappa (\mathbf{k})$, it satisfies criteria (i) and (ii) but not (iii). Instead, we take advantage of the $w_{m\mathbf{k}}$ smoothness in $\mathbf{k}$ and choose the following approximation in $w_{m\mathbf{k}}$ basis
\begin{equation}
\left\langle w_{m\mathbf{k+q}}\left| \text{sgn}(z-z_\kappa)e^{i\mathbf{q\cdot r}} \right|w_{n\mathbf{k}} \right\rangle  \approx \left[\tilde S_\kappa^{(W)}(\mathbf{k,q})\right]_{mn} \equiv \frac{1}{2}\left[ \tilde S_\kappa ^{(W)}(\mathbf{k}) + \tilde S_\kappa ^{(W)}(\mathbf{k+q}) \right].
\end{equation}
Finally, we arrive at the approximation which satisfies all these criteria
\begin{equation}
S_{\kappa mn}(\mathbf{k,Q}) \approx \left[ U_{\mathbf{k} + {\mathbf{q}}}\tilde S_\kappa ^{(W)}(\mathbf{k,q})U_\mathbf{k}^\dag \right]_{mn}.
\end{equation}
For $M_{\kappa mn}(\mathbf{k,q})$, we have the orthonormal relation $\left[ \tilde M_\kappa ^{(W)}(\mathbf{k,q}) \right]_{mn}=\delta_{mn}$ which leads to
\begin{equation}
M_{\kappa mn}(\mathbf{k,q}) \approx \left[ U_{\mathbf{k+q}}U_\mathbf{k}^\dag \right]_{mn}.
\end{equation}
This approximation for overlap matrix has already been used in the 3D \textit{ab initio} Fr{\"o}hlich model.

\begin{figure}
\includegraphics[width=8.6cm]{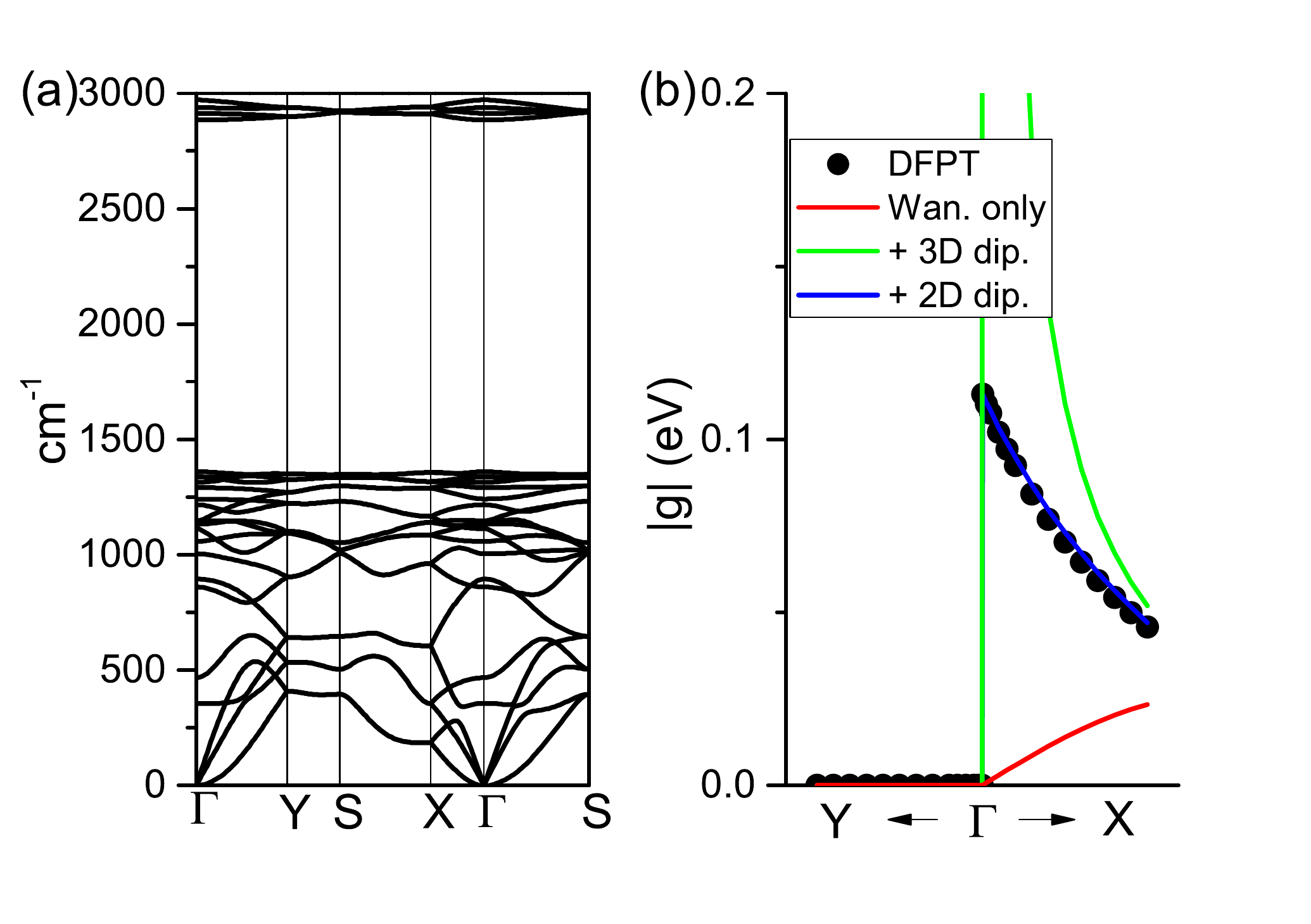}
\label{fig8}
\caption{(a) Phonon dispersion of boat-like graphane $(\text{C}_4\text{H}_4)_n$. (b) EPI between valence band maximum and the lowest optical phonon, which shows strong anisotropy and dipolar character. Only the 2D dipolar correction correctly reproduced the DFPT results.}
\end{figure}

In addition to the calculations shown in Figure 4, we also performed benchmark calculation again DFPT results for boat-like graphane, as shown in Figure 8. Again the 2D dipolar correction agrees well with the DFPT results while Wannier interpolation without or with 3D dipolar correction fails to reproduce the long-wavelength behavior.

To include the $\mathbf{e}^{\mathbf{-}\left| \mathbf{Q} \right|\left| \mathbf{z}\mathbf{-}\mathbf{z}_{\mathbf{\kappa}} \right|}$ factor in the Wannier Fourier interpolation, we need to treat it separately. Here we take $ S _{\kappa mn}(\mathbf{k,Q})$ as an example. If we assume the $w_{m\mathbf{k}}$ basis is complete, then we have $\sum_{m\mathbf{k}}\left| w_{m\mathbf{k}} \right\rangle\left\langle w_{m\mathbf{k}} \right| = 1$
which leads to
\begin{eqnarray}
&&\left\langle w_{m\mathbf{k + q}} \middle| \text{sgn}\left( z - z_{\kappa} \right)e^{i\mathbf{q}\mathbf{\cdot}\mathbf{r}}e^{- \left| \mathbf{Q} \right|\left| z - z_{\kappa} \right|} \middle| w_{n\mathbf{k}} \right\rangle \nonumber\\
&=& \sum_{m^{'}\mathbf{k}^{\mathbf{'}}}{\left\langle w_{m\mathbf{k + q}} \middle| \text{sgn}\left( z - z_{\kappa} \right)e^{i\mathbf{q}\mathbf{\cdot}\mathbf{r}} \middle| w_{m^{'}\mathbf{k}^{\mathbf{'}}} \right\rangle\left\langle w_{m^{'}\mathbf{k}^{\mathbf{'}}} \middle| e^{- \left| \mathbf{Q} \right|\left| z - z_{\kappa} \right|} \middle| w_{n\mathbf{k}} \right\rangle}.
\end{eqnarray}
When $\mathbf{k}$ and $\mathbf{k}^{'}$ are different, the last integration can be neglected due to different phase. By using $\sum_{m\mathbf{k}}\left| \left. \ w_{m\mathbf{k}} \right\rangle\left\langle w_{m\mathbf{k}} \right.\  \right| = 1$
and small $\mathbf{Q}$ limit we can further write
\begin{equation}
\left\langle w_{m^{'}\mathbf{k}} \middle| e^{- \left| \mathbf{Q} \right|\left| z - z_{\kappa} \right|} \middle| w_{n\mathbf{k}} \right\rangle = \left\{ \exp\left\lbrack - \left| \mathbf{Q} \right|d_{\kappa} \right\rbrack \right\}_{m^{'}n},
\end{equation}
\begin{equation}
\left\lbrack d_{\kappa} \right\rbrack_{m^{'}n} = \left\langle w_{m^{'}\mathbf{k}} \middle| \left| z - z_{\kappa} \right| \middle| w_{n\mathbf{k}} \right\rangle.
\end{equation}
Then we have
\begin{equation}
\left\langle w_{m\mathbf{k + q}} \middle| \text{sgn}\left( z - z_{\kappa} \right)e^{i\mathbf{q}\mathbf{\cdot}\mathbf{r}}e^{- \left| \mathbf{Q} \right|\left| z - z_{\kappa} \right|} \middle| w_{n\mathbf{k}} \right\rangle \approx \left\lbrack {\widetilde{S}}_{\kappa}^{\left( W \right)}\left( \mathbf{k,q} \right)\exp\left\lbrack - \left| \mathbf{Q} \right|d_{\kappa} \right\rbrack \right\rbrack_{{mn}}.
\end{equation}
If the matrix $d_{\kappa}$ is diagonalizable through unitary transformation $d_{\kappa} = V\Lambda V^{- 1}$ with $\Lambda$ being diagonal, the matrix exponential can be computed simply through
\begin{equation}
\exp\left\lbrack - \left| \mathbf{Q} \right|d_{\kappa} \right\rbrack = V\exp\left( -\left| \mathbf{Q} \right|\Lambda \right)V^{- 1}
\end{equation}
However, such process involves additional calculations, and using a relatively small Wannier basis may severely violate the completeness relation $\sum_{m\mathbf{k}}\left| w_{m\mathbf{k}} \right\rangle\left\langle w_{m\mathbf{k}} \right| = 1$. This calculation could therefore significantly increase the computational cost, thus losing the efficiency advantage. So, in this work, we choose to simply take the $e^{- \left| \mathbf{Q} \right|\left| z - z_{\kappa} \right|} \rightarrow 1$ approximation instead.

\section{Impact of perpendicular momentum sampling on 3D dipolar correction}
\begin{figure}
\includegraphics[width=8.6cm]{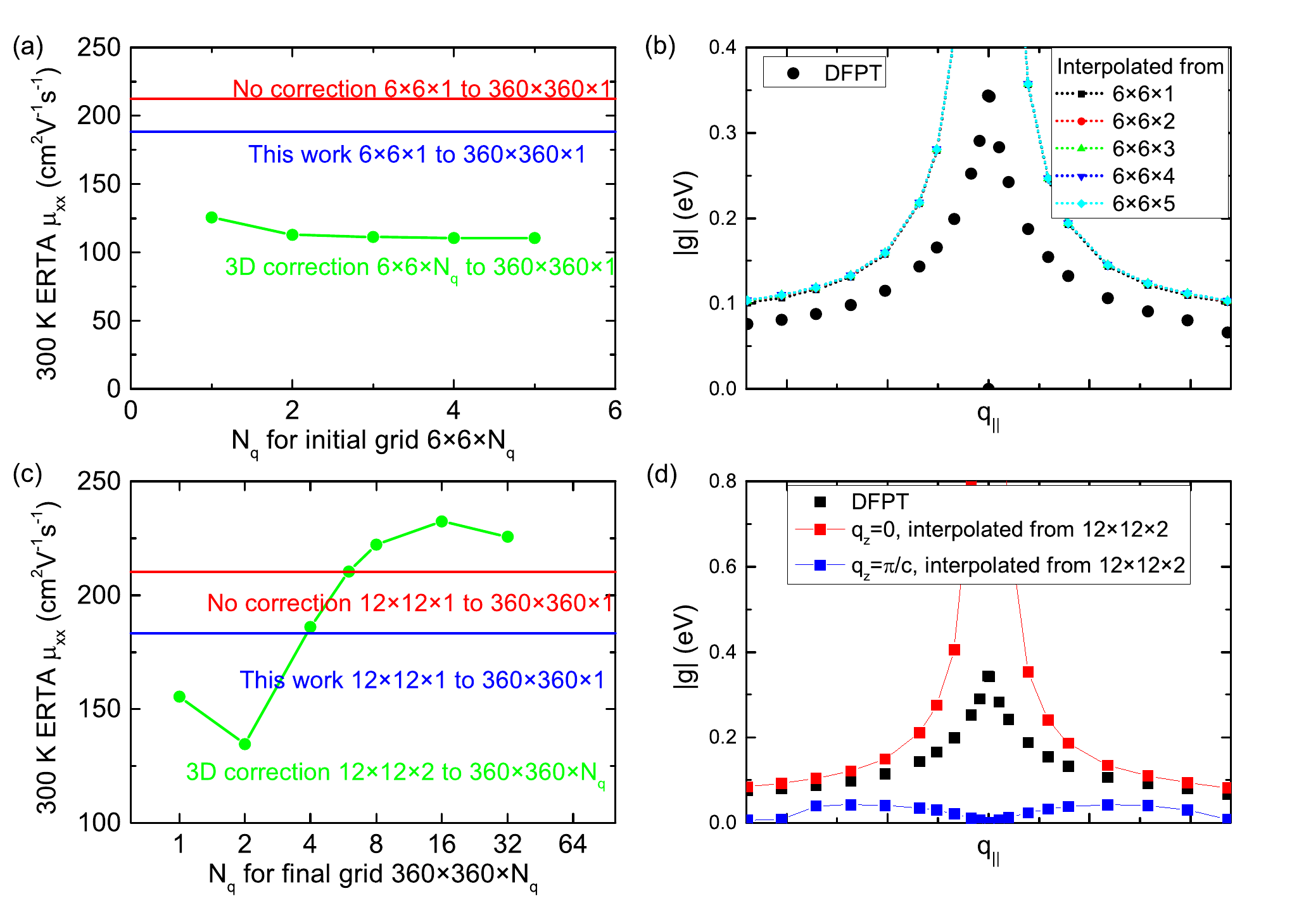}
\label{fig9}
\caption{Interpolation using 3D dipolar corrections combined with perpendicular momentum sampling. (a) $\text{MoS}_2$ electron mobility at 300 K from energy relaxation time approximation (ERTA) by interpolation from $6\times 6\times N_q$ with $N_q$ being 1 to 5. (b) Electron-phonon matrix element $|g|$ around $q=0$ interpolated from $6\times 6\times N_q$ grid. (c) $\text{MoS}_2$ electron mobility at 300 K computed with interpolation from $12\times 12\times 2$ to $360\times 360\times N_q$ with $N_q$ being 1 to 32. (d) Electron-phonon matrix element $|g|$ around $q_\parallel =0$ with different perpendicular component $q_z$.}
\end{figure}
We computed the electron mobility of $\text{MoS}_2$ at 300 K from energy relaxation time approximation (ERTA) by interpolation from $6\times 6\times N_{q}$ with $N_{q}$ being 1 to 5, as shown in Figure 9(a). The results are compared with calculations without dipolar correction (red) and with 2D correction in this work (blue). Interestingly, as shown in Figure 9, results are not improved with increasing $N_q$ and are already converged with $N_{q}\geq 2$. To understand this, we computed the interpolated electron-phonon matrix element $g$ for LO phonon around $q=0$ for all $N_q$ here. As shown in Figure 9(b), There is no significant change in long-wavelength $g$ with increased sampling. This is because for LO phonon, the interaction is dominated by dipolar effect as shown in Figure 1, while the short-ranged part has vanishing contribution. Therefore, even with increased sampling $N_q$, it still behaves as $g\propto 1/|q|$ at small $q$ limit and is not sensitive to the coarse grid sampling.

With this observation in mind, coarse grid with two q points along c-axis are used for the calculations below, which were also used by Li et al \cite{Li2019} and Ma et al \cite{Ma2020} in studying 2D InSe. We then performed convergence test for perpendicular sampling in the interpolated fine grid, starting from a denser $12\times 12\times 2$ coarse q-grid. Although the error becomes smaller than $5\%$ with $360\times 360\times 32$ grid, the result does not show convergence to the result with 2D correction, as shown in Figure 9(c). Instead, they become even higher than those without dipolar correction which should have already been incorrectly overestimated.

To understand the trend of mobility with 3D correction in Figure 9(c), we computed the interpolated electron-phonon matrix element $g$ for LO phonon around $q_\parallel =0$. They are computed for both $q_z=0$ (zone center) and $q_z=\pi/c$ (zone border), as shown in Figure 9(d). Interestingly, the $q_z=0$ results diverge while $q_z=\pi/c$ results vanish when $q_\parallel \to 0$. This is because for $q_z=0$, it still retains the 3D dipolar interaction behavior; while for the zone-border $q_z=\pi/c$, this is equivalent to partially removing the dipolar interaction due to out-of-phase cancellation, as has been discussed by Sohier et al in the case of phonon dispersion \cite{Sohier2017}. Therefore even though neither of the $q$-points reproduces the DFPT $g$, the overestimation due to 3D dipolar correction is partially remedied after integrating over $q_z$. This could explain why using additional $q_z$ sampling could improve the mobility prediction in certain cases \cite{Ma2020}. However, such improvement may not be guaranteed because it still cannot accurately reproduce the long-wavelength electron-phonon interaction, as we have shown in the case of $\text{MoS}_2$.

\bibliography{citations}{}

\end{document}